\documentclass[fleqn,usenatbib]{mnras}

\usepackage[T1]{fontenc}
\usepackage{ae,aecompl}

\usepackage{graphicx}	% Including figure files
\usepackage{amsmath}	% Advanced maths commands
\usepackage{amssymb}	% Extra maths symbols

\title{A dynamics-free lower bound on the mass of our galaxy}

\author[Zaritsky and Courtois]{
Dennis Zaritsky$^{1}$\thanks{E-mail: dennis.zaritsky@gmail.com}
and Helene Courtois$^{2}$
\\
$^{1}$Steward Observatory, University of Arizona, Tucson, AZ 85719, USA\\
$^{2}$Institut de Physique Nucl\'eaire, Universit\'e Lyon 1, CNRS/IN2P3, F-6922 Lyon, France
}

\date{Accepted November, 9, 2016}

\pubyear{2016}

\begin{document}
\label{firstpage}
\pagerange{\pageref{firstpage}--\pageref{lastpage}}
\maketitle

\begin{abstract}
We use a sample of Milky Way (MW) analogs for which we have stellar and disk gas mass measurements, published measurements of halo gas masses of the MW and of similar galaxies, and the well-established value of the cosmological baryon fraction to place a lower bound on the mass of the Galaxy of $7.7\times10^{11} M_\odot$ and estimate that the mass is likely to be $\ge 1.2\times10^{12}$ M$_\odot$. Although most dynamical analyses yield measurements consistent with these results, several recent studies have advocated for a total mass well below $10^{12} M_\odot$. We reject such low mass estimates because they imply a Galactic baryon matter fraction significantly above the universal value. Convergence between dynamical mass estimates and those based on the baryonic mass is an important milestone in our understanding of galaxies.
\end{abstract}


%%%%%%%%%%%%%%%%%%%%%%%%%%%%%%%%%%%%%%%%%%%%%%%%%%

%%%%%%%%%%%%%%%%% BODY OF PAPER %%%%%%%%%%%%%%%%%%

\section{Introduction}

An accurate measurement of the mass of our galaxy, the Milky Way (MW), is necessary for us to answer questions regarding the evolution of the Galaxy and its satellites, and to place the Galaxy within the larger cosmological context. Simple questions, such as whether the Magellanic Clouds are gravitationally bound to the Galaxy, remain unresolved given the full range of  published mass estimates for the MW \citep{besla, ka}. 

The direct way to estimate the mass of the MW, or any other galaxy, is to measure the kinematics of test particles, such as stars, globular clusters, and satellite galaxies, and to construct dynamical models that give rise to the observed kinematics. For nearly 30 years, since \cite{z89}, we have had kinematic measurements of test particles that probe the anticipated virial volume of the MW ($R \sim 200$ kpc). Although the estimated total mass resulting from that study ($> 10^{12}M_\odot$) remains consistent with a broad set of subsequent studies \citep{wilkinson,sakamoto,xue,li,watkins,mcmillan,bk,barber,kafie,eadie,huang}, some recent studies continue to advocate for significantly lower masses enclosed within 200 kpc ($< 4 \times 10^{11} M_\odot$ \citep{sofue09}, $(6.8\pm 4.1)\times 10^{11} M_\odot$ \citep{bhatt}, and $(5.6 \pm 1.2) \times 10^{11} M_\odot$ \citep{gibbons}). 
Even among studies that advocate for a total mass $> 10^{12} M_\odot$, their results vary from about $10^{12} M_\odot$ \citep{xue,mcmillan,huang} to well above $2 \times 10^{12} M_\odot$ \citep{sakamoto, li}. Why has the field not converged in 30 years?

Although the idea of dynamical modelling is simple in principle, it is complicated in practice for various reasons. First, the results can be strongly model dependent. In some cases, the dynamical tracers do not extend out to the virial radius. As such, the estimated virial mass is a model-based extrapolation of constraints on the enclosed mass at smaller radii \citep{kalberla,xue,huang}. Even without extrapolation, models can produce subtly different satellite populations than expected \citep{barber} subverting standard model assumptions. Second, we typically have incomplete kinematic information, usually only radial velocities for the test particles, and therefore uncertain orbits. This ambiguity translates to uncertainty in the derived mass of a factor of a few and can also lead to internal inconsistencies \citep[see, for example, discussion by][]{wilkinson}. 
Third,  the number of test particles, particularly at the largest distances, is small. This challenge has led some investigators to develop increasingly sophisticated statistical analyses with the aim of extracting robust estimates and uncertainties \citep{little,kochanek,eadie}. However, such efforts are compromised by their reliance on smooth, analytic models rather than the more realistic, cosmologically accurate ones. This shortcoming affects not only the nature of the dark matter halo, but also the characteristics of the satellites. 
For example, recent observations hint that some sets of satellite galaxies are not independent test particles but rather fell into the Galaxy halo as a bound group \citep{koposov}. Fourth, even if the individual satellites provide independent probes of the Galaxy mass, they are not a random set of tracer particles. At large radii, their orbital phases are non-random \citep{z89,white,bk}, and therefore no simple dynamical model in isolation will accurately reproduce the dependencies between Galaxy mass, galactocentric radius, and tracer particle velocity. 

It is telling that within the last 30 years, despite the discovery of numerous distant tracer particles, proper motion measurements, and better models, both the typical MW mass estimate and the uncertainty, as judged by the range of estimates themselves, have remained unchanged. This behaviour suggests that we are facing systematic uncertainties rather than random ones. The bottleneck will eventually be broken by superior kinematic data and models, but
in the interim, we propose a different approach, independent of dynamics, to provide competitive constraints on the lower bound of MW mass estimates.

The advent of precision cosmology provides new opportunities for the study of old problems. For our purpose, the relevant quantity is the ratio of the baryon density to the matter density, $f_b \equiv \Omega_{b}/\Omega_{m}$, or the baryon matter fraction.
 The availability of a precise determination of this value allows us to translate a measured total baryon mass to a total mass, assuming that galaxies have the cosmological baryon matter fraction. Because there may be baryonic mass loss during galaxy evolution \citep{larson}, our adoption of $f_b$ as appropriate for galaxies is conservative in that it biases downward the total mass lower bound.
 The \cite{planck} place $f_b$ at 0.155, while \cite{wmap} using data from the {\sl Wilkinson Microwave Anisotropy Probe} place it at 0.167 (WMAP7). The difference is minor for our purposes, consistent within their quoted uncertainties, and we adopt the {\sl WMAP} value.

It is difficult to complete a baryon inventory of our galaxy given our location within it. The effort requires substantial modelling of the observations \citep{flynn,mcmillan}. Even with the best of measurements, such a result represents a single example and leaves questions about the degree of possible variance among galaxies unanswered.  To address these difficulties,  we examine a sample of disk galaxies with the same rotation speed as the MW, which we refer to as MW analogs.

Given the baryons that we observe in these galaxies, what is the minimum total mass inferred if we adopt the cosmological baryon matter fraction, $f_b$? Using the mass distribution of the analogs, we find and describe below a bound defined by the 10th percentile ($7.7 \times 10^{11} M_\odot$) that is in significant tension with a subset of existing dynamical mass estimates and our preferred value ($> 1.2\times 10^{12}$ M$_\odot$) provides compelling support for the larger published mass estimates.
 In \S \ref{sec:data} we describe the data used in the baryon inventory. In \S \ref{sec:mass} we describe how we reach our mass estimates using a sample of galaxies similar to the MW and discuss various aspects of the estimate, including the role of systematic errors and where the most significant progress can be made. Where needed,
we adopt the simplified standard cosmology of $\Omega_m = 0.3$, $\Omega_\Lambda = 0.7$, and H$_0 = 70$ km s$^{-1}$ Mpc$^{-1}$.

\section{Data}
\label{sec:data}

The initial step in
our baryon accounting consists of summing cold gas mass estimates for galaxy disks, from H{\small I} fluxes, and
stellar mass estimates, from infrared magnitudes. The sum of these two components is {only an initial step because we will not have yet included baryons in the halo}. We will address this shortcoming in \S \ref{sec:mass} using published studies.

\subsection{Cold disk gas masses}

We draw our H{\small I} data from the Cosmic Flows project, which has gathered and consistently remeasured digital H{\small I} spectra from the public archives of the largest radio-telescopes worldwide.
Tens of thousands of galaxy line widths were measured or remeasured using a new robust method described by \citet{courtois09,courtois11}. A detailed description of the measurement and corrections for relativistic broadening, broadening due to finite spectral resolution, and internal turbulent motions are presented and discussed \citet{courtois09,courtois11} and \cite{tully12}.  We have already used these data in previous papers regarding the baryonic Tully-Fisher relation \citep{btf} and unusual galaxies \citep{courtois15}.

We currently have coherent H{\small I} measurements for 13,213 galaxies. This catalog is available for public use at the Extragalactic Distance Database (EDD) website\footnote{http://edd.ifa.hawaii.edu; catalog ``All Digital H{\small I}"} and we call it the ``All Digital H{\small I} catalog".
Several other parameters available are included \citep{tully09} such as the integrated H{\small I} line fluxes computed from the H{\small I} lines, which have a flux calibration uncertainty of about 10 to 15\%, and the average heliocentric velocities. We use the smooth Hubble flow distances, calculated from the CMB-frame recessional velocities. 

Of interest here from that same dataset is $W_{m50,i}$, which is a measure of the width of the H{\small I} 21 cm emission line, corrected for inclination, and hence corresponding roughly to twice the disk rotation velocity. An important distinction among Tully-Fisher studies is that this measurement is not necessarily either the peak rotation velocity nor the asymptotic rotation velocity of a flat rotation curve. Measurements of the rotation obtained from resolved rotation curves can lead to lower-scatter, presumably more accurate versions of the scaling relation \citep[for example, see][]{mcgaugh}. The scaling relation is not the focus here and we use the rotation velocities only to group galaxies as similar.

A key aspect, and source of uncertainty, in the measurement of $W_{m50,i}$ is the disk inclination, $i$, used to correct the rotation velocity for projection. The inclinations are obtained from the Hyperleda data based on measurements from optical images.  To mitigate errors, we limit the sample to highly inclined galaxies ($i > 65^\circ$) for which the projection corrections are then small. Increasing the inclination limit to 75$^\circ$ reduces the size of the sample but does not affect our conclusions.
 
To calculate cold gas disk masses, we use the H{\small I} flux from the database using 
$M_{\rm HI} =2.36 \times 10^5  D_L^2  F,$
where $M_{\rm HI}$ is the H{\small I} gas mass, $D_L$ is the luminosity distance in Mpc, and $F$ is the flux integrated within the H{\small I} line profile in units of Jy km s$^{-1}$.
To obtain an estimate of the full gas mass, rather than just H{\small I}, we 
us a sliding correction scale for $M_{\rm H_2}$ with galaxy type \citep{mcg}, and correct for the mass in He and metals by multiplying by 1.4.

We only consider galaxies with recessional velocities $>$ 2000 km s$^{-1}$ to minimise distance errors due to peculiar velocities and morphological types Sab or later to focus on galaxies where rotational support dominates. 

\subsection{Stellar masses}

We estimate the stellar masses using 3.6$\mu$m fluxes, obtained either with the {\sl Spitzer} Space Telescope \citep{werner} or WISE \citep{wright}. To convert IR fluxes to stellar masses, we use the calibration based on the analysis by \cite{mcgaugh}, who calculate disk, stellar mass-to-light ratios, $M/L$'s, by requiring consistent Tully-Fisher relations for gas-rich and star-rich galaxies. They conclude that $M/L = 0.45$ in solar units at 3.6$\mu$m, with an intrinsic  scatter among galaxies that is less than 30\%. 

This result is entirely consistent with an independent determination of the stellar $M/L$ at 3.6$\mu$m done
using the resolved stellar populations in the Large Magellanic Cloud \citep{eskew}. 
The conversion advocated by \cite{eskew} is $M_* = 10^{5.97} F_{3.6} ({D \over 0.05})^2$, where $F_{3.6}$ is in Jy, $M_*$ in solar masses, and $D$ is the distance of the source in Mpc. For $M_{3.6, \odot} = 3.24$ \citep{oh} and a 3.6$\mu$m zero point of 280.9 Jy \citep{reach}. The \cite{eskew} calibration is equivalent to adopting $M/L$ of 0.53 in solar units at 3.6$\mu$m. 
A relevant factor in that calculation is that it adopts a Salpeter initial mass function \citep{salpeter} to correct for stars below the detection limit.  As such, the resulting stellar masses are biased high if the alternative Kroupa \citep{kroupa} or Chabrier \citep{chabrier} mass functions are in fact the correct ones. This discrepancy may be the reason why the \cite{eskew} calibration results in stellar masses that are 18\% larger, although such values are still within the \cite{mcgaugh} limit on the intrinsic scatter. 

\section{The Mass of the Galaxy}
\label{sec:mass}

\begin{figure}
\includegraphics[width=\columnwidth]{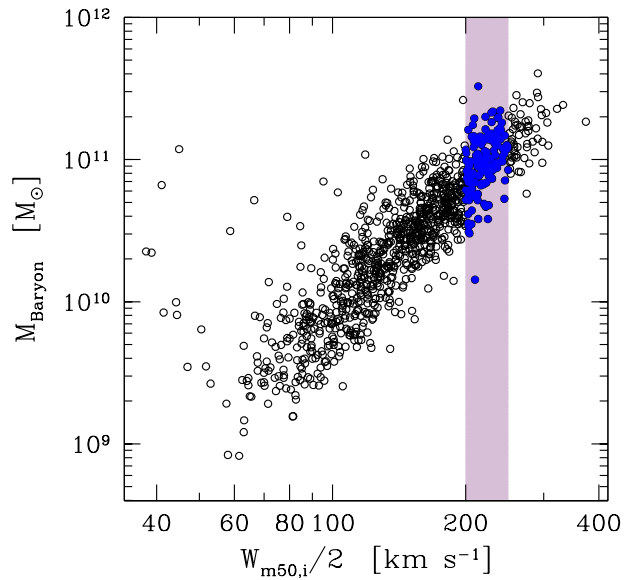}
\caption{The relationship between baryonic mass of galaxies (stars plus disk cold gas only) relative to the width of the inclination-corrected H{\small I} emission line, $W_{m50,i}/2$. Our sample of galaxies with $200 \le W_{m50,i}/2  \le 250$ km s$^{-1}$ is highlighted with the filled blue circles and vertical shaded band.}
\label{fig:btf}
\end{figure}

In Figure \ref{fig:btf} we plot the baryonic mass accounted for by the combination of the cold disk gas and stars that we have traced as described above as a function of $W_{m50,i}/2$. The existence of the evident correlation is what gives rise to the baryonic Tully-Fisher relation. 

There is no direct observation of $W_{m50,i}$ for the MW. There are, however, numerous measurements of the rotation curve from well inside the solar radius to about twice the solar radius \citep{burton,gunn,clemens,fich,brand,honma,olling,sofue,huang}. Although the details depend on the adopted values of the LSR Galactocentric distance and velocity, these studies place the bulk of the material between a few kpc radius to twice the solar radius at circulate velocities between 200 and 250 km s$^{-1}$. The one among these studies that advocates for possibly lower circular velocities \citep{olling} obtains those estimates using a Galactic centre distance that  is significantly smaller than the currently accepted value \citep{reid}. As such, we adopt galaxies with rotation widths, $W_{m50,i}/2$, between 200 to 250 km s$^{-1}$ as representative of the MW. $W_{m50,i}$ is not necessarily the rotation speed at the solar radius, nor is it the ``flat" asymptotic value of the rotation curve. $W_{m50,i}/2$ measures the maximum rotation speed of a substantial amount of the H{\small I} in the galaxy, and so is an (upwardly) biased measurement of the rotation curve. With the existence of a number of studies that place much of the Galaxy's gas at rotation speeds as high as 250 km s$^{-1}$,  including galaxies with $W_{m50,i}/2$ as high as 250 km s$^{-1}$ is appropriate. 

\begin{figure}
	\includegraphics[width=\columnwidth]{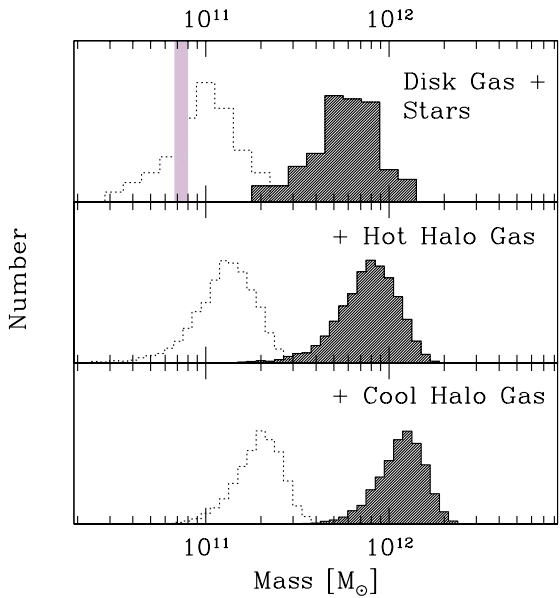}
    \caption{Mass distributions. We present the baryonic mass distributions for the MW analog sample, dotted lines, and the projected total masses, shaded histograms. Each panel represents a step along our baryon accounting, beginning at the top where we include only the disk gas and stars. The vertical shaded region represents the combined, independent estimates of the masses of these components for the MW. The next two panels show the results when we add the hot halo gas (middle panel) and when we then subsequently add the cool halo gas (bottom panel).}
    \label{fig:mass}
\end{figure}

We have 151 galaxies that are highlighted in Figure \ref{fig:btf} and constitute our sample of MW analogs. One is possibly an outlier that lies too far above the bulk of the sample, and one that lies too far below. We remove these from the sample. Otherwise, the data seem representative of the galaxy sample as a whole in that they follow the mean trend of baryon mass with $W_{m50,i}/2$ and show a typical degree of scatter.
We present the resulting distribution baryon masses for the set of analogs in Figure \ref{fig:mass}. The upper panel of that Figure includes only the components discussed so far,  disk gas and stars. 

The baryon accounting is difficult in our own galaxy but there have been attempts and these provide a basic test of our results. Combining the published stellar disk mass \citep[$(6.43 \pm 0.63) \times 10^{10} M_\odot$;][]{mcmillan} and disk gas mass
 \citep[($9.5 \pm 0.3)\times 10^{9} M_\odot$;][]{dame} places the MW slightly below the median
 of the sample, but within 1$\sigma$ of the mean (Figure \ref{fig:mass}). 

The summation of the mass of the stars and disk gas within these galaxies is known to be an incomplete baryon accounting because of the existence of significant amounts of gas in galaxy halos, including our own \citep{bregman07,gupta, putman}.
A grossly incomplete baryon accounting will lead to an underestimate of the total mass. Even so, the median total mass determined from only these baryons, 
$5.1\times10^{11}M_\odot$ (see Table \ref{tab:masses}), is already larger than some of the dynamical total MW mass estimates. For the baryon inventory to be consistent with the smallest dynamical mass estimates in the literature, such as those of \cite{sofue09} and \cite{gibbons}, our analog galaxies would have to have their full cosmological allotment of baryons (no baryonic mass loss) and have all of those baryons locked within their disks (no additional baryons in their halos).

The second of these two conditions can already be ruled out. The MW does have a significant amount of gas in its halo. Regarding the 
hot (T $> 10^5$ K) gas, we refer to recent measurements of that component in the Galaxy \citep{gupta,miller,nicastro}. 
These studies agree that there is substantial mass in hot halo gas, $> 10^{10}$ M$_\odot$, although the derived values vary depending on the model assumptions, even within in a single study \citep[from $2^{+3}_{-1}\times 10^{10}$ to $1.3^{+2.1}_{-0.7}\times10^{11}$ M$_\odot$;][]{nicastro} . 
We adopt an intermediate value, consistent within the uncertainties of both extremes $4.3^{+0.9}_{-0.8} \times 10^{10} M_\odot$ \citep{miller}, with a modification that we increase the uncertainty to $3\times10^{10}$ M$_\odot$ to allow for the lower values among the range of 
estimates.

Adding this gas to our accounting, sampling 10,000 times from a Gaussian distribution defined by the adopted mean and revised $1\sigma$ uncertainty, we obtain the baryon and total mass distributions shown in the middle panel of Figure \ref{fig:mass}. The corresponding mass constraints are given in Table \ref{tab:masses} and consist of a median total mass of $7.9\times10^{11} M_\odot$ and a 10th percentile lower bound of $4.7 \times 10^{11} M_\odot$. We are beginning to be able to statistically rule out the smallest published estimates of the MW mass \citep[eg.][]{sofue}.

Accounting for the cool
(T $\sim 10^4$ K) halo gas is more complicated because we must rely on measurements of this component in MW analogs rather than in the MW halo itself. 
By examining and modelling the metal ion absorption lines within the halos of $L_*$ galaxies, \cite{werk} place a lower limit on the mass of such gas at $6.5 \times 10^{10} M_\odot$ per galaxy halo. Because the average galaxy in their sample has a slightly lower stellar mass than in ours (log $M_*$ of 10.6 for their sample in comparison to 10.87 for our sample of analogs and 10.81 for the MW\citep{mcmillan}), their lower limit is likely to also be a lower limit for our sample and for the MW. There are complex questions regarding the physical nature of this gas and its relation with the hot gaseous halo \citep{werk16}, but the ubiquitous existence of this component has recently been confirmed through the measurement of the H$\alpha$ emission profile out to 100 kpc projected radius in galaxy halos using over 7 million SDSS sightlines \citep{zhang}.

Formally \cite{werk} claim their mass estimate to be a lower limit, suggesting that we should add the full amount to all of our analog galaxies. However, in the interest of capturing some level of uncertainty in this difficult measurement, we adopt a 1$\sigma$ uncertainty of 50\%. Again we draw from a Gaussian, this time with a mean of $6.5 \times 10^{10} M_\odot$ and an uncertainty of $3.25 \times 10^{10} M_\odot$. Adding this contribution, we find the distribution of masses shown in the lowest panel of Figure \ref{fig:mass}. The corresponding mass constraints are again presented in Table \ref{tab:masses} and correspond to a median mass of $1.2\times 10^{12}$ M$_\odot$ and a 10th percentile lower bound of $7.7\times10^{11}$ M$_\odot$.

How can one escape these conclusions? We see only two potential, but unattractive, paths. First, the contribution from the cool halo gas is not confirmed for the MW. Although we know of no reason why the MW would be distinct from other L$_*$ galaxies in this particular way, the measurement we use is not a direct measurement of this component in the halo of the MW. 
If one is willing to postulate the absence, or gross overestimation, of the cool gas in the MW halo, then the lower bound could drop to as far as $4.7\times10^{11}$ M$_\odot$. However, an independent argument against dismissing these halo baryons comes from matching numerical galaxy simulations to the baryonic Tully-Fisher relation, which requires that $\sim$ 40\% of a galaxy's baryons are in the disk \citep{btf}. That argument provides no guidance on the phase of the halo baryons, but the estimate that $\sim$ 60\% of the baryons are in the halo is consistent with the sum of the masses independently measured for the cool and hot halo phases that we cite above. Second, if $W_{m50,i}$ for the MW is lower than the range we adopted (from 200 to 250 km s$^{-1}$) then the stellar and cold disk gas masses would correspondingly decrease (see Figure \ref{fig:btf}). Again, we see little indication that this is a viable possibility given previous empirical results on the MW rotation curve. 

If we accept the results of the baryon accounting presented at face value and the typical dynamical total mass estimate of $\sim 10^{12}$ M$_\odot$, we find that the community is 
converging on several important results. First, we understand at a broad level the constituent baryonic components of galaxies like the Milky Way. Conjectured possibilities of large populations of dark objects (halo white dwarfs, black holes, dense cold molecular gas clouds) cannot contribute significantly to the total baryon budget. Second, galaxies like the Milky Way do not lose significant fractions ($\gg$ 10\%) of their baryons to the intergalactic medium. Third, the disks of such galaxies contain at most $\sim$ 40\% of the all the baryons initially within the halo.
Confirming these conclusions depends primarily on improving the constraints on the mass of the hot and cold gaseous components of galaxy halos and in developing better upper limits on the dynamical mass estimates. The latter is generally not an area of focus because the dark matter problem has driven a focus on determining lower rather than upper bounds on the halo masses.

\begin{table*}
\caption{Critical Assumptions and Adoptions}
\label{tab:summary}
\begin{tabular}{lll}
\hline
Category&Value&Comment\\
\hline
cosmological baryon matter fraction&$0.167\pm0.012$&WMAP7 \citep{wmap},  Planck results lead to larger total mass estimates\\
halo baryon fraction&cosmological&no expectation that it be above the cosmological value, unaccounted\\
&&baryonic mass loss from galaxies will cause us to underestimate total masses\\

3.6$\mu$m stellar mass-to-light ratio&$0.45\pm0.14$ &from \cite{mcgaugh}, range includes \citet{eskew},\\
&(M$_\odot$/L$_\odot$)& calibration, uncertainty is roughly consistent with IMFs ranging from\\
&& \cite{salpeter} to \cite{kroupa}\\
MW circular velocity [km sec$^{-1}$]&$220^{+30}_{-20}$ km sec$^{-1}$&from various studies cited in text\\
cool gas mass&$> 6.5\times 10^{10}$ M$_\odot$&not specifically for MW, but a lower limit for slightly lower stellar mass galaxies\\
&& \citep{werk}, we treat as mean value and add an uncertainty of 50\%\\
hot gas mass &4.3$^{+0.9}_{-0.8}\times 10^{10}$ M$_\odot$ &specifically for MW \citep{miller}, but we increase the uncertainty to \\&&$3\times10^{10}$ M$_\odot$ to span other estimates \citep{nicastro}\\

\hline
\end{tabular}
\end{table*}

\begin{table*}
\caption{MW Mass Constraints}
\label{tab:masses}
\begin{tabular}{lll}
\hline
Distribution &Median&10th Percentile\\
 &[M$_\odot$]&[M$_\odot$]\\
\hline
baryons only, disk gas + stars&$9.8\times10^{10}$&$3.5\times 10^{10}$\\
total mass using disk gas + stars&$5.1\times10^{11}$&$2.7\times10^{11}$\\
total mass using disk \& hot halo gas + stars&$7.9\times10^{11}$&$4.7\times10^{11}$\\
total mass using disk, cool \& hot halo gas + stars&$1.2\times10^{12}$&$7.7\times10^{11}$\\
\hline
\end{tabular}
\end{table*}

\section{Conclusions}

With the ever improving census of baryons in 
galaxies similar to the MW, and in the MW itself, we are at a point where converting the baryon mass into an estimate of the total mass using the well-measured cosmological baryon matter fraction yields constraints on the total mass that are competitive with dynamical estimates.
We find that to the degree we have a complete baryon census, the MW must have a total mass $>7.7 \times 10^{11}$ M$_\odot$.
We expect this to be a slightly conservative lower bound primarily because 1) galaxies are expected to have less than the cosmological baryon matter fraction due to mass loss arising from energetic feedback from stellar evolution and/or central engines \citep{larson,opp}, and 2) our baryon inventory is still likely to be incomplete. The largest source of uncertainty in this estimate arises in the accounting of the MW cool halo gas.

The use of the baryon fraction to set bounds on the mass of the Galaxy results in a limit that is in conflict with several recent dynamical studies advocating masses well below $10^{12}$ M$_\odot$ \citep{sofue09, bhatt, gibbons}. However, it must be noted that dynamical mass estimates have bounced around for nearly the past 30 years, without much sign of convergence. Our dynamics-free lower bound on the mass of the Galaxy provides an independent way that may help guide us to convergence on this issue. Ever improving baryonic inventories, particularly a measurement of the cool MW halo, may provide the strictest lower bounds on the total mass of the MW and other galaxies for some time.
Once we reach convergence between the dynamical mass estimates and the total mass estimates derived from the baryon mass, we will be able to close several long-standing questions regarding galaxies.

\section*{Acknowledgements}

DZ thanks the Universit\'e of Lyon 1 for financial support during a collaborative visit there that resulted in this paper and to everyone there for their welcoming friendship. DZ also acknowledges financial support from NSF AST-1311326 and the University of Arizona.

%%%%%%%%%%%%%%%%%%%%%%%%%%%%%%%%%%%%%%%%%%%%%%%%%%



%\bibliographystyle{mnras}


%%%%%%%%%%%%%%%%%%%%%%%%%%%%%%%%%%%%%%%%%%%%%%%%%%

% Don't change these lines
\bsp	% typesetting comment
\label{lastpage}
\end{document}